\documentclass{aastex61}

\submitjournal{ApJ}

\shorttitle{Semantic Metadata for Astronomy and Astrophysics}
\shortauthors{Frey, K. \& Accomazzi, A.}

\begin{document}

\title{The Unified Astronomy Thesaurus: Semantic Metadata for Astronomy and Astrophysics}

\correspondingauthor{Katie Frey}
\email{kfrey@cfa.harvard.edu}

\author[0000-0001-9891-44657]{Katie Frey}
\affiliation{Wolbach Library, Harvard-Smithsonian Center for Astrophysics \\
60 Garden Street \\
Cambridge, MA 02138, USA}

\author[0000-0002-4110-3511]{Alberto Accomazzi}
\affiliation{Astrophysics Data System, Harvard Smithsonian Center for Astrophysics \\
60 Garden Street \\
Cambridge, MA 02138, USA}

\begin{abstract}
Several different controlled vocabularies have been developed and used by the astronomical community, each designed to serve a specific need and a specific group.  The Unified Astronomy Thesaurus (UAT) attempts to provide a highly structured controlled vocabulary that will be relevant and useful across the entire discipline, regardless of content or platform.  As two major use cases for the UAT include classifying articles and data, we examine the UAT in comparison with the Astronomical Subject Keywords used by major publications and the JWST Science Keywords used by STScI's Astronomer's Proposal Tool.
\end{abstract}




\section{Introduction}

Astronomy research relies on a network of distributed, heterogeneous data archives which provide observations, measurements, articles, software, and services to the scientific community.  The number of existing scholarly documents has been doubling every 15 years for centuries \citep{deSolla61}.  The growth rate in research data has been much greater, and its observed doubling time of one year for the past decade will further decrease in the coming era of LSST\footnote{\url{https://lsst.org}} and SKA.\footnote{\url{https://www.skatelescope.org/}}  Together, these create significant challenges for scientists who must explore and understand a much larger body of knowledge in order to make contributions to their field.  As the amount and complexity of this data has increased, the community has successfully adopted standards and protocols which have facilitated access and analysis of data retrieved from these archives \citep{2010ivoa.rept.1123A}.  Building tools and standards to support the discoverability of research material is a critical step remaining in this process.  From reviewing recent publications on a particular class of astronomical objects, to finding datasets with high-resolution observations of those objects, to retrieving code which implements appropriate data reduction techniques, having the ability to locate literature, data, code, and supporting material is a prerequisite to conducting research.

The need for semantic-aware data services has long been recognized by information professionals working on text analysis and bibliographic information retrieval \citep{salton83,miller} as well as those involved in bibliographic services in astronomy \citep{1992PASAu..10..134S,2000A&AS..143...85A}. With the emergence of the Virtual Observatory and the rise of Astroinformatics as a new discipline \citep{2009astro2010P...6B}, this need becomes even more pressing \citep{2008ASPC..394..742G,2009ASPC..411..179G,2009ASPC..411..165G,2011ASPC..442..415A}.  As an example, if an astronomer searched an astronomy data archive for "compact objects," she would expect results related to black holes and binary systems.  To accomplish even the simplest level of a semantic-aware search (i.e., one which can leverage a list of synonyms) requires the formal encoding of underlying disciplinary knowledge.  The shared adoption by the astronomy community of systems and technologies supporting the formal organization of concepts, resources, and their inter-relationships is an essential pillar of this infrastructure effort.

In this paper, we describe a community-led effort to codify the most common concepts used in modern astronomy into a single thesaurus, called the Unified Astronomy Thesaurus\footnote{\url{http://astrothesaurus.org/}} (UAT), a knowledge organization system which can be used to describe the expressions of concepts and their organizational structure in a formal way.  In the next section, we give an overview of the history behind the Thesaurus and discuss the principles behind its design and development.  Section 3 compares the Thesaurus to the traditional keyword system used in astronomical publications and to a keyword system used to describe observing proposals in a major NASA archive.  Section 4 discusses the current and envisioned uses of the Thesaurus, listing some of its early adopters.

\section{A New Thesaurus}
\subsection{History}

The Unified Astronomy Thesaurus is the most recent iteration of efforts to develop a keyword classification system that can be used to describe the astronomical concepts and entities found within literature, software, conferences, proposals, datasets, and other astronomical resources.  The International Astronomical Union (IAU) Thesaurus,\footnote{\url{http://www.mso.anu.edu.au/library/thesaurus/}} compiled by Robyn Shobbrook of the Australian Astronomical Observatory and Robert Shobbrook of the University of Sydney in 1992, is often considered the first such attempt \citep{1992PASAu..10..134S}. The IAU Thesaurus was last updated in 1995 and contained 2,551 terms organized into a loose hierarchy.  In 2008, an update to the IAU Thesaurus was created by Rick Hessman and labeled as the International Virtual Observatory Alliance, or IVOA, Thesaurus.\footnote{\url{http://www.astro.physik.uni-goettingen.de/~hessman/rdf/IVOAT/index.html}}  Hessman's work added about 300 terms, bringing his proposed IVOA Thesaurus to 2,889 terms.
 
Around the time of the initial development of the IAU Thesaurus, editors of the major astronomy and astrophysics journals worked together to create an organized set of keywords for classifying and indexing journal articles.  This system is called the Astronomical Subject Keywords\footnote{\url{http://journals.aas.org/authors/keywords2013.html}} (ASK) and was adopted for use in 1992 \citep{2014ASPC..485..461A}.  Although the limitations of this classification scheme have become apparent in the intervening years, this set of terms is still currently in use.  The Astronomical Subject Keywords were last updated in 2013, and contain about 370 concepts organized by general subject area.

The Physics and Astronomy Classification Scheme\footnote{\url{https://publishing.aip.org/publishing/pacs/pacs-2010-regular-edition}} (PACS) is a classification scheme containing roughly 400 astronomical terms, developed by the American Institute of Physics (AIP) in the 1970s to annotate the physics literature using a hierarchical set of codes.  PACS was proposed in 1975, updated every two years, and served as the main organizational scheme for the major physics journals until it was decommissioned in 2011 in favor of developing a new system that would support modern indexing and searching technology.
 
In 2010, major publishers in physics and astronomy, together with other stakeholders from the IVOA, the NASA Astrophysics Data System (ADS), and the Harvard-Smithsonian Center for Astrophysics Library began to explore new approaches to the various classification schemes in use.  Specifically, the Institute of Physics (IoP) and AIP decided to collaborate, funding a project to merge the astronomy portion of their vocabularies with the enhanced IVOA Thesaurus to create a more robust thesaurus for astronomy.  The resulting work was named the Unified Astronomy Thesaurus and was donated to the American Astronomical Society (AAS), who hold ownership and copyright.  In order to further develop and promote the UAT, the AAS established a Steering Committee, which currently includes members from the AAS, the ADS, the library community, and other astrophysics institutions.

\subsection{Design}
The UAT has been designed as an open, interoperable, and community-supported project that embraces modern standards for thesaurus curation and linked data formats.  It uses methods and tools that make it easy for members of the astronomical community to suggest term additions, refinements, revisions, and deletions, allowing open contributions for the maintenance of the thesaurus.  The thesaurus is currently distributed as a SKOS\footnote{\url{https://www.w3.org/2004/02/skos/}} (Simple Knowledge Organization System) document, a standard format that uses RDF (Resource Description Framework) for structured controlled vocabularies.  This will allow for the UAT to be integrated into a variety of applications, acting as a bridge between disparate systems.

Although there are a range of options for organizing a controlled vocabulary---all the way from a flat list of unconnected concepts to a structured arrangement of precise relationships---the format of a thesaurus was determined to best support the goals of search and findability.  According to the ANSI/NISO standard on controlled vocabularies, a thesaurus is a "controlled vocabulary arranged in a known order and structured so that the various relationships among terms are displayed clearly and identified by standardized relationship indicators" \citep{ansiniso}.  This allows us to define some general relationships between concepts using standard SKOS properties (e.g. "broader," "narrower," "related," "preferred label," and "alternate label") while avoiding overly complex structures.

Linking concepts together in child/parent relationships allows a user to follow a path from broader terms to narrower terms and to discover more specific terms to assist in finding the concepts that best describe their resource.  This structure also helps to give additional context to each term, which can help clarify it's meaning.  More importantly, this hierarchy can inform searching and indexing applications that child terms are subsets of their parent terms, and can be used to suggest related, alternative results to the user.

The UAT has been licensed under a Creative Commons Attribution-ShareAlike license (CC-BY-SA).  This license was chosen by considering both the nature of the Thesaurus as a community-supported free cultural work and the intended process for its use and development.  The CC-BY-SA license grants anyone the freedom to copy, use, or improve the work while requiring that derivative works be made available under the same license.  This license reflects the collaborative process which has led us to the creation and steady improvement of the UAT and safeguards its future openness.  Incidentally, having a license for the UAT was recognized as a necessity early on: the lack of licensing terms for the progenitors of the UAT has been a cause of concern for several publishers interested in reusing these works in their production systems.

\subsection{Scope}

Early on, it was decided that the UAT should contain all the terms necessary to completely describe the primary topics found in the mainstream refereed literature of astronomy and astrophysics.  As the discipline expands over time, so, too, will the thesaurus; for instance, additional terms describing exoplanets and gravitational waves will surely need to be incorporated into the Thesaurus within the next few years.  In addition to deciding what should be included in the UAT, the boundaries of the Thesaurus have been intentionally delineated to define what must be excluded.  Without a clearly articulated definition of scope, the task of maintaining a vocabulary can quickly get out of control as additional terms are incorporated.  Concepts that are only tangentially connected to astronomy, or methods and techniques which are not specific to astronomy, are better described in external vocabularies, which can be referenced by, rather than incorporated into, the UAT.  Similarly, existing taxonomies that detail astronomical objects, instruments, and facilities are best kept outside of the UAT and managed by existing groups who are currently curating them.

The curatorial decisions limiting which concepts are represented in the UAT and which are kept out do not necessarily mean that descriptiveness needs to be compromised for the sake of expedience: other thesauri can provide the concepts needed to properly describe a multidisciplinary work, one which could never be fully described using an astronomy-focused thesaurus.  One of the features of the Semantic Web \citep{SWEB} is the ability to interlink distributed concept schemes, and SKOS provides ways to formally map, re-use, or even delegate entire branches of a thesaurus to an external knowledge systems.

Concept mapping provides a way to establish semantic similarity between a term in the UAT and an external, potentially broader, knowledge base.  Using concept mapping allows a curator for a project such as WikiData \citep{WIKIDATA} to link concepts in Wikipedia to corresponding concepts in the UAT, stating, for instance, that the UAT concept for "G stars" is equivalent to the Wikipedia category "G-type star."  As more resources are linked to WikiData, such curation allows for cross-vocabulary mapping, which will increase interoperability across knowledge systems and disciplines.  This is the approach behind the Linked Data effort \citep{LOD}, which attempts to publish and interlink structured data on the web.  While this is, in general, a good thing, one should be aware of the fact that any such mapping is subject to interpretation and approximation, which means that the result of these efforts may present new challenges to those attempting to leverage them.

Re-using a concept allows a system to "borrow" it from an external source via a simple referencing mechanism.  A publisher could use this mechanism to say that the top-level UAT concepts represent the astronomy branch of a higher-level physics vocabulary, which could also include high energy physics concepts to fully describe an interdisciplinary cosmology paper.  Finally, the SKOS standard allows one to extend a thesaurus by delegating parts of it to outside knowledge systems rather than incorporating their contents.  One useful application of this capability for the UAT could be the delegation of object types to the SIMBAD ontology \citep{2010ivoa.rept.0303D}.  This delegation works both ways, and allows a publisher such as IoP to simply adopt the UAT itself as the astronomy branch of its physics thesaurus.  It should be noted that the choice of relying on external knowledge systems involves making certain compromises, the most notable of which is a delegation of authority, and therefore control, over part of the domain being covered.

\section{Comparisons with Existing Vocabularies}

A major goal of the Unified Astronomy Thesaurus is to be used as the keyword indexing scheme for major publications in the field of astronomy, such as the \textit{Astrophysical Journal},  and for major astronomical dataset collections, such as the ones curated by the Mikulski Archive for Space Telescopes (MAST).  Towards that goal, we want to ensure that the UAT compares favorably with systems already in use: the Astronomical Subject Keywords\footnote{\url{http://journals.aas.org/authors/keywords2013.html}} and the Science Categories and Keywords\footnote{\url{https://jwst-docs.stsci.edu/display/JPPOM/JWST+Scientific+Keywords}} for the Astronomer's Proposal Tool developed for MAST.  In order to accomplish this, we analyzed how the UAT compared with these two controlled vocabularies.

The goal of this exercise was to investigate how well UAT concepts could be used to represent the terms in existing vocabularies and whether the structure of the UAT was appropriate to convey the relationships between them. These analyses were completed by human experts, rather than performed automatically with any sort of machine learning tool.   Without experts to define the relationships between concepts, a computer cannot accurately discern the true meaning of a concept and decide if something slightly broader or narrower truly requires the addition of a new concept.  For example, without a curated vocabulary, a computer coming across the words "quasi-stellar radio sources" and "quasar" would not be able to tell that the two words refer to the same concept.  On the other hand, concepts that are very similar syntactically may have wildly different meanings, such as "gravity waves" and "gravitational waves."  Only someone with subject knowledge in the field can make the connections and distinctions, defining the relationships (or lack thereof) that a computer could later rely upon.  Though these manual analyses require a higher time commitment, each additional mapping and analysis project helps to inform strengths and weakness of the Unified Astronomy Thesaurus, thereby building a better tool.

\subsection{Journal Editor's Astronomical Subject Keywords}

The concepts found in the Astronomical Subject Keywords (ASK) were approved and adopted by the editors of major astronomy journals such as the \textit{Astronomical Journal}, \textit{Astronomy \& Astrophysics}, \textit{Monthly Notices of the Royal Astronomical Society}, the \textit{Publications of the Astronomical Society of the Pacific}, and the \textit{Astrophysical Journal} (including \textit{Letters} and \textit{Supplements}).  The Keywords were originally introduced in 1973 by the \textit{Astrophysical Journal} under the name of "Subject Headings" \citep{1972BAAS....4..402.} and were later adopted by the other journals.  Revisions and updates to the Astronomical Subject Keywords were made by general agreement and consensus among the editors of the above journals every few years based on emailed suggestions.  The last of these updates was in 2013, after which a general consensus emerged that these keywords were not as useful as originally envisioned. 

Throughout these revisions, the Astronomical Subject Keywords remained a list of a relatively small number of concepts organized into broad categories.  Furthermore, the ASK is a simple text list, utilizing none of the modern standards and formats to support machine readability and usability.  Innovations in semantic markup and knowledge representation have brought about standards such as SKOS.  Systems that follow these standards can leverage these relationships and use them to build organizational structures that benefit from the explicit links.  For example, a system tasked to search for very specific leaf concepts at the bottom of the hierarchy, such as "UAT: Bailey type stars," might suggest that a user broaden their results by searching for the concept's parent, "UAT: RR Lyrae variable stars." 
 
In order to facilitate a transition from the Astronomical Subject Keywords to the UAT and as a way to validate the completeness of the latter, we decided to attempt to map ASK terms to UAT concepts.  For each term in the Astronomical Subject Keywords, we searched for similar terms found in the Unified Astronomy Thesaurus.  To perform the mapping, we used well-known SKOS Mapping Properties\footnote{\url{https://www.w3.org/TR/skos-reference/##mapping}} to describe the relationship between ASK terms and UAT concepts, as defined in Table 1.

\begin{deluxetable}{l|l}
\tablenum{1}
\tablecaption{Definition of SKOS Mapping Properties used to describe the relationship between concepts}
\tablehead{
\colhead{SKOS Mapping Property} & \colhead{Definition}}
\startdata
exactMatch & the term in one vocabulary describes exactly the same concept in the other vocabulary \\
closeMatch & the term in one vocabulary closely describes a similar concept in the other vocabulary \\
narrowMatch & the term in one vocabulary is a specific example or narrower instance of the similar \\[-6pt]
 & concept(s) described in the other vocabulary \\
broadMatch & the term in one vocabulary is a more general version or broader instance of the \\[-6pt]
 & similar concept(s) described in the other vocabulary \\
relatedMatch & the term in one vocabulary describes a concept similar but not the same as the concept(s) \\[-6pt]
 & described in the other vocabulary \\
\enddata
\end{deluxetable}

The analysis defined 557 relationships between terms in the Astronomical Subject Keywords and concepts in the Unified Astronomy Thesaurus: 187 exact matches, 9 close matches, 265 narrow matches, and 96 related matches.  We also found that 62 terms in the ASK had no analogues to concepts in the UAT.
 
Overall, terms in the Astronomical Subject Keywords tend to describe broader concepts or multiple concepts, whereas the Unified Astronomy Thesaurus uses a single term to describe a single, specific concept.  For example, the single term "ASK: meteorites, meteors, meteoroids" describes three separate UAT concepts ("UAT: Meteorites," "UAT: Meteors," and "UAT: Meteoroids").  Similarly, "ASK: accretion, accretion disk" describes both the process of accretion and the generic concept of an accretion disk, and several UAT concepts are a narrow match such as "UAT: Accretion," "UAT: Galaxy accretion," and "UAT: Stellar accretion disk," among others.  Further examples of relationship matches between ASK and UAT can be seen in Table 2.

\begin{deluxetable}{l|c|l}
\tablenum{2}
\tablecaption{Some examples of concept relationships between ASK and UAT.  This table is read thusly: "intergalactic medium has narrow match cool intergalactic medium."}
\tablehead{
\colhead{ASK term} & \colhead{SKOS Mapping Property} & \colhead{UAT concept}}
\startdata
(galaxies:) intergalactic medium & exacMatch & Intergalactic medium \\
(galaxies:) intergalactic medium & narrowMatch & Cool intergalactic medium \\
(galaxies:) intergalactic medium & narrowMatch & Hot intergalactic medium \\
(galaxies:) intergalactic medium & narrowMatch & Warm-hot intergalactic medium \\
(galaxies:) intergalactic medium & relatedMatch & Intergalactic medium phases \\
(galaxies:) intergalactic medium & relatedMatch & Intergalactic gas \\
(ISM:) cosmic rays & exactMatch & Cosmic rays \\
(ISM:) cosmic rays & narrowMatch & Galactic cosmic rays \\
(ISM:) cosmic rays & relatedMatch & Cosmic ray astronomy \\
meteorites, meteors, meteoroids & narrowMatch & Meteorites \\
meteorites, meteors, meteoroids & narrowMatch & Meteors \\
meteorites, meteors, meteoroids & narrowMatch & Meteoroids \\
methods: analytical & relatedMatch & Analytical mathematics \\
techniques: photometric & closeMatch & Photometry \\
methods: data analysis & (no match) & \nodata \\
\enddata
\end{deluxetable}

The high number of close, partial, and narrow matches, along with the 96 related concept links between the ASK and the UAT, implies two things:
\begin{enumerate}
\item The UAT has a lack of generalized concepts and
\item For the most part, terms in the ASK are included as concepts in the UAT.
\end{enumerate}

Where the ASK has only top-level categories and second level concepts, the UAT has 10 levels of depth.  Likewise, the UAT has 1,843 concepts where the ASK has only 302 terms.  It is apparent that the UAT delves into more topics, and covers them in greater detail, than does the ASK.  The exception to this rule is that the ASK allows for users to include specifically named astronomical objects.  For example, if an author wished to use the keyword "ASK: comets: individual (..., ...) they would be expected to include the individually named comet in parentheses, i.e. "ASK: comets: individual (Shoemaker–Levy 9)."  In contrast to ASK, and in accordance with best practices on controlled vocabularies \citep{ansiniso}, the UAT does not allow for open-ended, on-demand user input and only includes a few specifically named examples.

The lack of open-ended and generalized concepts in the UAT is a feature rather than a bug.  That being said, this comparison highlights 62 terms in the ASK that have no match of any kind with concepts in the UAT.  For example, where the ASK includes concepts describing different facets of publishing (such as "ASK: editorials, notices" and "ASK: errata, addenda"), there are no corresponding terms in the UAT.  However, they could easily fit into the existing UAT section on "Astronomical research."  Similarly, the ASK covers modern topics on mathematical and computational methods for processing data and images much more thoroughly than the UAT, which includes only a few very general concepts on these topics.  Whether these concepts warrant inclusion in the UAT or not will depend on their relevance to astronomy literature and the existence of external SKOS vocabularies which the UAT could link to that cover these concepts in greater depth.

Finally, though the UAT has many concepts about various physical processes, they tend to be specific and collocated with astronomical phenomena.  In contrast, the ASK allows these concepts to stand alone so that they can be discussed in general rather than only as they relate to specific objects or events.  This difference leads to an important practical distinction between the two systems: terms in the ASK may be ambiguous and their meaning can only be inferred when considered alongside other terms assigned to the same resource.  Instead, the UAT uses precise language to define each concept so that its meaning does not depend on its context.  As an example, consider the term "ASK: accretion, accretion disks," which appears under the category of "Physical Data and Processes."  When considering this term alone, there is no way for a reader to know whether it refers to stellar, galactic, or Bondi accretion processes, which correspond to very different concepts in the UAT.  While a human reader may rather quickly find the answer to the question, things are more complex for automated systems which will use this information to classify content and provide recommendations based on them.  Since one of the goals of the UAT is to be machine actionable, the greater precision of its semantics is a significant advantage over the ASK.

\subsection{JWST Scientific Keywords}

With the imminent launch of the James Webb Space Telescope (JWST), the Space Telescope Science Institute (STScI) has begun the work of updating and preparing the Astronomer's Proposal Tool, which is used to write, validate, and submit proposals.  A specific change under consideration for this work is to update the two keyword lists used to describe proposals and targets within the Proposal Tool.  One keyword list, the Target Descriptions,\footnote{\url{https://jwst-docs.stsci.edu/display/JPPOM/Target+Descriptions}} has already been updated with a subset of UAT keywords and will be used to index JWST data going forward.  The other list under consideration to be updated is the JWST Scientific Keywords,\footnote{\url{https://jwst-docs.stsci.edu/display/JPPOM/JWST+Scientific+Keywords}} which is currently based on the keywords used to describe proposals for the Hubble Space Telescope.

The JWST Scientific Keywords (JSK) list currently stands at about 140 concepts arranged into seven top-level categories and second level keywords.  MAST collaborators recently mapped the JSK terms to the UAT in order to investigate the potential of aligning the two.  The analysis completed by STScI focused on the viability of using the UAT with the Astronomer's Proposal Tool.  We found that nearly two-thirds of the JSK were well represented in the UAT, either as exact matches (approx. 31\%) or as synonymous matches (approx. 34\%), where the concepts had the same meaning but used a different term or syntax.  The remaining JSK terms were split between concepts missing entirely in the UAT (approx. 14\%), and concepts requiring further analysis to make a determination (approx. 21\%).  See Table 3 for an excerpt from this analysis.
 
The last group also contains JSK terms that described multiple concepts (e.g., "JSK: Bulges, Spheroids, and Ellipticals") which could easily match multiple concepts within the UAT if they were split into their component parts (e.g., "UAT: Galaxy bulges," "UAT: Dwarf spheroidal galaxies," and "UAT: Elliptical galaxies").  In some cases, the compound JSK terms still had no matches even when split into separate concepts.  Careful review of all the JSK terms with multiple matches in the UAT is recommended before making any further determinations of which terms might need updating in either the JSK or the UAT.

Much of the reason why additional review is required for concepts in the JSK is that the concepts do not have unique identifiers throughout the system, leading to semantic ambiguity.  The concept "JSK: Astrometry" is found under the categories "JSK: Stellar Physics" and "JSK: Stellar Populations," but in both cases, the meaning of "JSK: Astrometry" is clearly the same.  It could be said that Astrometry as found in the JSK terms simply has two parent concepts (also called a polyhierarchical relationship).  However, the concept of "Dust" stands in stark contrast to this: "JSK: Dust" is found under "JSK: Stellar Physics," "JSK: Galaxies and the IGM," and "JSK: Stellar Populations"---three different parent concepts!  To compound the issue, the concept of "Dust" clearly has different meanings relative to "Stellar Physics" vs. "Galaxies/IGM."  In the former case ("JSK: Stellar Physics, Dust"), it is equivalent to "UAT: Interstellar dust," but in the latter case ("JSK: Galaxies and the IGM, Dust"), it is clear that "UAT: Intergalactic dust clouds" is a more appropriate analogue.  As for the third case ("JSK: Stellar Populations, Dust"), it is unclear whether "Dust" shares a meaning with "UAT: Intergalactic dust clouds" or with "JSK: Interstellar dust." 

Just as with the ASK, this analysis uncovered that there are small but significant gaps in UAT content, even though the majority of JSK terms are well-represented.  This analysis demonstrated that the UAT lacks concepts regarding observational techniques, modern computational methods for analysis, and niche concepts relating to galaxies, galactic processes, and black holes.  At the same time, the comparison showed shortcomings in the current JSK system, where ambiguous terms are used to describe semantically different concepts.  While the context in which the terms appear explains the difference in their meaning, it is best practice not to have a vocabulary reuse the same term for different concepts.

\begin{longrotatetable}
\tablenum{3}
\begin{deluxetable*}{ccccc}
\tablecaption{Sample of Comparison of JWST Subject Keywords to Unified Astronomy Thesaurus \label{chartable}}
\tablewidth{700pt}
\tabletypesize{\scriptsize}
\tablehead{
\colhead{JWST Scientific Category} & \colhead{JWST Scientific Keyword} & 
\colhead{Match w/ UAT} & \colhead{UAT Equivalent} & 
\colhead{Suggested New UAT Concept}
} 
\startdata
Cosmology & Chemical Abundances & Exact match & Chemical abundances & \nodata \\
Cosmology & Clusters Of Galaxies & Equivalent & Galaxy clusters & \nodata \\
Cosmology & Cooling Flows & Review & \nodata & Cooling flows \\
Cosmology & First Light Stars And Galaxies & Multi-Match & Protostars & \nodata \\
Cosmology & First Light Stars And Galaxies & Multi-Match & Protogalaxies & \nodata \\
Cosmology & Supernovae & Exact match & Supernovae & \nodata \\
Galaxies and the IGM & Bulges, Spheroids, And Ellipticals & Multi-Match & Galaxy bulges & \nodata \\
Galaxies and the IGM & Bulges, Spheroids, And Ellipticals & Multi-Match & Dwarf spheroidal galaxies & \nodata \\
Galaxies and the IGM & Bulges, Spheroids, And Ellipticals & Multi-Match & Elliptical galaxies & \nodata \\
Massive Black Holes And Their Host Galaxies & AGN Host Galaxies & Not in UAT & \nodata & AGN host galaxies \\
Massive Black Holes And Their Host Galaxies & Jets & Equivalent & Stellar jets & \nodata \\
Massive Black Holes And Their Host Galaxies & Liners & Equivalent & LINER galaxies & \nodata \\
Massive Black Holes And Their Host Galaxies & M-Sigma Relation & Not in UAT & \nodata & M-sigma relation \\
Massive Black Holes And Their Host Galaxies & Quasars & Exact match & Quasars & \nodata \\
Massive Black Holes And Their Host Galaxies & Quenched Galaxies & Not in UAT & \nodata & Quenched galaxies \\
Planets and Planet Formation & Planets and Planet Formation & Equivalent & Exoplanet astronomy & \nodata \\
Planets and Planet Formation & Biomarkers & Not in UAT & \nodata & Biomarkers \\
Planets and Planet Formation & Extrasolar Planets & Equivalent & Exoplanets & \nodata \\
Planets and Planet Formation & Planetary Atmospheres & Equivalent & Exoplanet atmospheres & \nodata \\
Planets and Planet Formation & Planetary Satellites & Review & Exomoons & Natural satellites [Extrasolar] \\
Planets and Planet Formation & Space Weather & Review & \nodata & Space weather [Exoplanets] \\
Solar System & Solar System & Equivalent & Solar system astronomy & \nodata \\
Solar System & Biomarkers & Not in UAT & \nodata & Biomarkers \\
Solar System & Chemical Composition & Review & Astrochemistry & \nodata \\
Solar System & Planetary Atmospheres & Exact match & Planetary atmospheres & \nodata \\
Solar System & Planetary Satellites & Equivalent & Natural satellites & \nodata \\
Solar System & Space Weather & Review & \nodata & Space weather \\
Stellar Physics & Stellar Physics & Exact match & Stellar physics & \nodata \\
Stellar Physics & Atmospheres & Equivalent & Stellar atmospheres & \nodata \\
Stellar Physics & Binaries & Equivalent & Binary stars & \nodata \\
Stellar Physics & Brown Dwarfs & Equivalent & Brown dwarf stars & \nodata \\
Stellar Physics & Dust & Equivalent & Interstellar dust & \nodata \\
Stellar Populations & Chemical Abundances & Exact match & Chemical abundances & \nodata \\
Stellar Populations & Irregular Galaxies & Exact match & Irregular galaxies & \nodata \\
Stellar Populations & Local Group Galaxies & Equivalent & Local Group & \nodata \\
\enddata
\end{deluxetable*}
\end{longrotatetable}

\section{Looking Forward}

Following the comparison between the JSK and the UAT, the decision was made to align the keywords used to describe proposals for HST and JWST with the UAT through two methods.  First, changes will be made to JSK where appropriate to use the concepts found in the UAT.  Second, STScI will submit update requests to the UAT to increase coverage in areas such as infrared astronomy and exoplanetary research, thus benefiting the products of both communities.  Once this is done, the complete overhaul of the proposal keywords will be presented to HST and JWST user communities for feedback and final approval.  Revising the JSWT Scientific Keywords to match UAT concepts will ensure that the Astronomer's Proposal Toolkit is aligned with a standardized vocabulary, recognized in the wider astronomical community, which we hope will influence other observatories to adopt the UAT for similar purposes.

Other early adopters of the UAT include the ADS, the AAS, and the IVOA.  The ADS has been looking for keyword systems that can be used across its bibliographic collection to describe and classify documents, which can provide better ways for users to filter the literature and data for its recommendation engine.  Early tests have indicated that the UAT could be used in conjunction with text mining techniques to automatically extract a set of concepts discussed in a paper with reasonable accuracy.  As this process is further developed, we expect the results to become available in the ADS system, allowing users to select, narrow, or broaden search results based on one or more of the concepts discussed.  

Publishers such as the AAS are considering the use of concepts drawn from the UAT to describe papers being submitted to their journals in the future.  The workflow envisioned includes a semi-automated process where concepts are suggested during the editorial process and validated by the author or journal manager.  By incorporating this content during manuscript submission, publishers can ensure that their content is assigned high quality disciplinary metadata which can be used further downstream to support its discoverability and improve its classification and selection.  The UAT is also being adopted as the primary source of subject keywords for the Virtual Observatory Registry with the upcoming VOResource version 1.1.\footnote{\url{http://ivoa.net/documents/VOResource/20171107}}  Our hope is that further adoption by the community will follow similarly to what happened with the ASK in the mid-90s.

Long-term success of the UAT depends on continuous feedback from the community.  Beyond exhaustive comparison studies such as those described in this paper, there are a number of other methods for submitting feedback to the Unified Astronomy Thesaurus.  The most preferred way is by opening a new Issue on the UAT's GitHub page\footnote{\url{https://github.com/astrothesaurus/UAT}} with any comments or suggestions.  The GitHub Issues platform has proven to be an excellent method for tracking incoming suggestions while also allowing for open discussion.  GitHub is also the platform used to store current and past versions of the UAT, along with release notes.
 
A user can also submit feedback by using the UAT Sorting Tool.\footnote{\url{http://uat.altbibl.io/}}  This web-based JavaScript platform allows a user to explore the full hierarchy of the UAT and to submit suggestions to change the content of the UAT.  Suggestions for new concepts, concepts to be deprecated, and modifications to the organizational structure of the thesaurus can be visualized using the Sorting Tool.  By using the form found in the tool, all changes made in the visualization, along with any comments provided by the user, are automatically collected and submitted by email to the curators.
 
Recognizing that the curation of the UAT requires both an editorial process which may at times involve resolving conflicts and also regular maintenance to ensure a sound technical structure, the stakeholders endorsed the appointment of a Curator for the Unified Astronomy Thesaurus, an astronomy librarian charged with looking after the well-being of the thesaurus and promoting its proper growth.  Responsibilities of the UAT Curator include overseeing the overall curation process, ensuring the long-term stability of the UAT, maintaining versioned releases, moderating discussions within the editorial community, and managing outreach activities.  The UAT Curator receives all feedback and works with editors and researchers in astronomy to validate suggested changes, which are implemented in future releases of the Unified Astronomy Thesaurus.

We expect that the Thesaurus will be reviewed and potentially updated at least once a year to ensure that it is still current with respect to the latest research in the field. This schedule should allow enough time to analyze recent literature to detect whether any new concepts have appeared.  Community feedback and use cases may dictate a more frequent update cycle, which could be as frequent as every quarter.  In order to facilitate the deployment of a new version of the UAT, we adopted a versioning policy\footnote{\url{http://astrothesaurus.org/about/releases-versioning/}} modeled on Semantic Versioning\footnote{\url{https://semver.org/}} to allow users to consistently determine the scope of the change in the UAT and whether a version update will require changes in their applications and use cases.  Additionally, the Curator will release detailed patch notes specifying the changes and providing mappings between the previous and current version if necessary.

Astronomy and astrophysics are relatively fast-changing disciplines, where new theories and discoveries spawn new research fields on a regular basis.  Since one of the main goals of the UAT is to provide a formal language that can be used to describe this entire field, we need to ensure that the UAT remains a living document, regularly updated and revised to capture any new concepts introduced in the scholarly literature.  Comparisons with vocabularies currently in use by astronomers, such as the two described above, help to evaluate whether the UAT meets those goals, shed light on areas in need of improvement, and help to define its general scope.  Our hope is that these comparisons will lead to a continuous improvement of the UAT, which will then become the unifying thesaurus for our disciplines, providing an interoperable semantic layer across a variety of scholarly content (including proposals, datasets, software, literature, and objects) in astronomy and astrophysics.

\acknowledgments

Many thanks to Sarah Weissman and Jenny Novacescu at the Space Telescope Science Institute for their work on the JWST Science Keyword and Unified Astronomy Thesaurus comparison.

The UAT is ultimately based on the 1992 IAU Thesaurus of Shobbrook and Shobbrook, with subsequent additions from Helen Knudsen, Marlene Cummins and Liz Bryson, influence from the consensus list of journal keywords, and, in the late-2000s, updating work by Rick Hessman under the auspices of the IVOA.

We are thankful to the community of collaborators from around the world who have contributed to the UAT over the years, and in particular to those who provided a vision for it in the early days: Graham McCann, Norman Gray, Christopher Erdmann, and Chris Biemesderfer.

\end{document}